\documentclass[amsmath,amssymb, aps, prl, twocolumn]{revtex4-1}

\usepackage{graphicx}
\usepackage{dcolumn}
\usepackage{bm}
\usepackage[usenames]{color}

\begin{document}

\title{Unified non-Fermi liquid and superconductivity in a model of strongly-correlated systems}

\author{Yi Zhang}
\email{frankzhangyi@gmail.com}
\affiliation{Department of Physics, Cornell University, Ithaca, New York 14853, USA}
\affiliation{Kavli Institute for Theoretical Physics, University of California, Santa Barbara, California 93106,
USA}

\date{\today}

\begin{abstract}
We study a model of strongly-correlated systems that incorporates phases such as Fermi liquids, non-Fermi liquids, and superconductivity, in addition to potential intertwined orders. The model describes Fermi surfaces of spinful electron gas or electron liquid coupled to bosons via an intermediate gauge field. Effectively, the coupling imposes constraints and interactions between the fermion spin and the local boson density. This grants the bosons the meaning as the Schwinger boson of the magnetic order and allows us to probe a larger phase space, rather than around the quantum critical point. We design the initial model so that after the boson and gauge fields are integrated out exactly, the resulting fermion-only effective theory only consists of several local interactions, allowing controlled weak-coupling interpretation for certain parameter regions. Consequently, we find a non-Fermi liquid whose degrees of freedom in the effective theory take the form of an \emph{emergent} Fermi surface that has a different Luttinger volume from the original fermions, therefore explicitly violates the Luttinger theorem. Further, this allows a $d$-wave superconducting instability if the effective interacting induced by boson coupling is in the relevant pairing channel. When the emergent Fermi surface undergoes a Lifshitz transition, the carrier type changes. In addition, we attribute the Mott insulating behavior in the strong coupling limit to the loss of emergent Fermi surface and argue the possibility of coexisting magnetic order. Our effective theory also suggests the possibility of charge density waves as a consequence of \emph{effective} strong repulsion and is independent of the Fermi-surface-nesting scenario. We discuss the possible optimal conditions for higher superconducting transition temperature and potential relevances and implications for realistic materials.\end{abstract}
\maketitle

\emph{Introduction-- }The discoveries of high-temperature superconductors in Cu-based\cite{Bednorz1986, Hazen1988, Sheng1988, Sheng1988b, Schilling1993} and Fe-based\cite{Takahashi2008, Sasmal2008, Rotter2008, Ren2008, He2013, Liu2012, Ge2014} materials bring hope for room-temperature superconductors with countless promising applications. However, after about three decades of intensive research, the origin of high-temperature superconductivity is still not clear and remains highly controversial, hindering targeted engineering for enhanced Tc. The unlikelihood of an electron-phonon attraction mechanisms\cite{Leggett2006} that succeeded in conventional superconductivity calls for alternatives in unconventional pairing mechanism. The situation becomes further complicated due to the existence of intertwined orders\cite{KivelsonIntertwine} in the phase diagrams of superconductors such as cuprates, iron pnictides, and chalcogenides. Other than the proximate non-Fermi liquid, various other symmetry-breaking orders such as antiferromagnetism (AFM) state\cite{Lake2002}, spin density waves (SDWs)\cite{Ichikawa2000, Klauss2000, Lee2004, Udby2013}, charge density waves (CDWs)\cite{Hoffman2002, Kohsaka2007, Wise2008, Wu2011, Achkar2012, Ghiringhelli2012, Chang2012}, nematic order\cite{Ando2002, Chu2010, Daou2010, Lawler2010, Chu2012, Gallais2013} and others have been observed or conjectured, making a consistent description and unified microscopic theory difficult in dimensions large then one\cite{KivelsonIntertwine}. The rich possibilities that the critical fluctuations of these orders may induce or assist pairing have also brought various interesting yet controversial proposals\cite{Berg2012, Lederer2015, Lederer2017, Xu2017}. The confusion is further fueled by the lack of theory descriptions of non-Fermi liquids.

Experiments on hole-doped cuprate materials have also unveiled a series of mind-boggling properties. ARPES experiments have observed a full Fermi surface with consistent Luttinger volume in overdoped materials, yet only disconnected Fermi arcs in underdoped materials\cite{Damascelli2003}. In contrast, heat capacity measurements in underdoped materials are much smaller than the expectation\cite{Riggs2011}, and quantum oscillations in underdoped materials have also unveiled a much small pocket with increasing volume and diverging mass upon increased doping\cite{Ramshaw2015}, therefore unusually contradicting the APRES observations. Recently, Hall coefficient measurements observed a drastic difference between the carrier densities in underdoped and overdoped materials\cite{Badoux2016, Balakirev2003}. In addition, there are both materials from the Cu-based\cite{SATO1997, Bozovic2002} and Fe-based\cite{He2013, Liu2012, Ge2014} families, whose monolayers and thin films have a comparable or even higher transition temperature than the bulk materials. Importantly, while the quantum criticality of antiferromagnetism may have an important role in driving the superconductivity, the onset and especially the optimization of the superconductivity is not adjacent to the demise of the antiferromagnetic order.

Inspired by the richness of the neighboring phases and phenomena, we propose and study in this paper a model of strongly-correlated systems that can, for the first time, unitedly incorporate Fermi liquids, non-Fermi liquids, Mott insulators, and superconductivity, as well as possibilities of magnetic and charge density wave orders. The model describes a weak-coupling or non-interacting Fermi surface of spinful electrons coupled to bosons via an intermediate gauge field. Effectively, the gauge constraints project the boson numbers to the fermionic spins and give them interpretations as the Schwinger bosons of the magnetic order. Unlike previous studies that focus around the quantum critical point\cite{Berg2012}, our setup allows us to probe a larger phase space from the Fermi fluid to the possible coexisting magnetic order with the Mott insulator. We find the model illuminating since the boson and gauge fields can be integrated out analytically and exactly, and the resulting fermion-only effective theory only has a limited number of local interactions. In certain parameter regions, these interactions are weak enough to allow a controlled interpretation of the effective theory and offer insights into the low-energy physics of the original strong-coupling theory.

In the weak-coupling phase space in the effective theory, there exist scenarios where the Luttinger volume of the Fermi surface differs from that of the original fermions, therefore explicitly violating the Luttinger theorem. Consequently, we establish a non-Fermi liquid whose degrees of freedom take the form of an \emph{emergent} Fermi surface. Sometimes, the emergent Fermi surface undergoes a Lifshitz transition and the carrier type changes. In the parameter region where both an emergent Fermi surface and effective interaction in the relevant pairing channel emerge, the system is unstable towards superconductivity. The $d$-wave superconducting gap symmetry is preferable due to the nearest-neighbor nature of the interactions.

When the impact of the bosons to the Fermi surface gets even stronger, the Fermi surface in the effective model can shrink further and vanish completely, corresponding to a Mott insulator driven by interaction instead of filling. The simultaneous Mott insulating phase of the Schwinger bosons suggests that there may also be coexisting magnetic order. Also, we attribute the possibility of charge density wave order to effective strong repulsion induced by boson renormalization, which is independent of the weak-coupling scenario involving Fermi surface nesting. In addition, we conjecture the optimal conditions for enhanced superconducting transition temperature. This model is \emph{sign-problem-free}. Therefore our predictions and conjectures can be verified with Monte Carlo calculations. In the main text, we mainly focus on the theoretical model itself for its phase diagram and properties. At the end, we also discuss the potential relevance and implications to realistic materials.

\emph{Initial model of interacting fermions and bosons-- }We consider strongly correlated systems with bosonic fields $\theta$ coupled to fermionic fields $\chi$ in a $2+1d$ space-time lattice:
\begin{equation}
Z=\int DbZ_{\theta}[b]Z_{\chi}\left[-b\right]
\label{eq:totalaction}
\end{equation}
where $b$ is an internal $U(1)$ gauge field. The boson action is given by:
\begin{eqnarray}
Z_{\theta}\left[b\right]&=&\int D\theta e^{-\mathcal{L}
_{\theta}\left[b\right]} \nonumber\\
-\mathcal{L}_{\theta}\left[b\right]&=& \sum_{n\mu} -h\left(\theta_{n+\hat{\mu}}-\theta_{n}-b_{n\mu}\right)\nonumber\\
&=&\sum_{n\mu}\ln\left[1+2\alpha\cos\left(\theta_{n+\hat{\mu}}-\theta_{n}-b_{n\mu}\right)\right.\nonumber\\
&+&\left.2\beta\cos\left(2\theta_{n+\hat{\mu}}-2\theta_{n}-2b_{n\mu}\right)+\cdots\right]
\label{eq:baction}
\end{eqnarray}
where $\alpha$ and $\beta$ are dimensionless quantities for single and double hopping of rotor bosons between nearest neighbors, respectively; $\cdots$ denotes the higher order terms, and $h\left(\theta\right)=h\left(\theta+2\pi\right)$ is a real function. As an example, for $XY$-coupling $h\left(\theta\right)=-J\cos\theta$ we have $\alpha=I_1(J)/I_0(J)$ and $\beta=I_2(J)/I_0(J)$, where $I_n$ is the modified Bessel function of the first kind of $n$th order. Smaller $\alpha$ and $\beta$ prefers a bosonic Mott insulator while larger $\alpha$ and $\beta$ prefers a superfluid. In particular, $\alpha, \beta\rightarrow 1$, i.e. $J\rightarrow \infty$ in the above example, indicates the superfluid at the zero-temperature limit. $\alpha$ and $\beta$ are our major tuning parameters on the boson dynamics.

\begin{figure}
\includegraphics[scale=0.4]{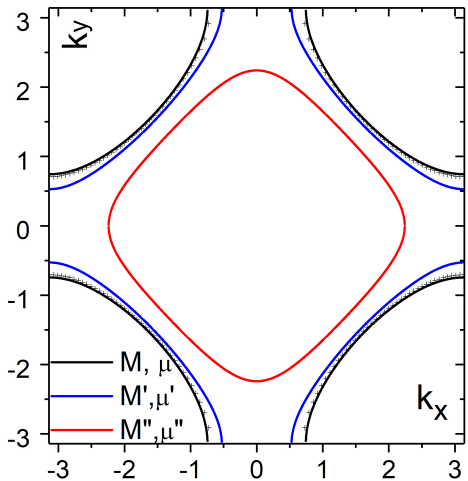}
\caption{The Fermi surfaces of the non-interaction electrons in Eq. \ref{eq:caction} and Eq. \ref{eq:cham}. Black: $M=4$ and $\mu=3.33$ gives a hole pocket; blue: $M'=5$ and $\mu'=4.17$ also gives an hole pocket yet with slightly modified Luttinger volume; red: $M"=8$ and $\mu"=6.67$ gives an electron pocket instead; the cross symbols almost overlapping with the black curves are for $\tilde M=4.2$ and $\tilde \mu=3.5$. If we start with the non-interacting electrons parameterized by $M$ and $\mu$ and couple them to bosons with parameter $\alpha$, the resulting fermion quadratic terms in the effective theory are parameterized by $M/\alpha$ and $\mu/\alpha$ instead after the bosons are integrated out. See later discussions in the main text.  Therefore, the black curves, the cross symbols, the blue curves, and the red curve, can be regarded as the \emph{emergent} Fermi surfaces after no renormalization $\alpha=1$, weak renormalization $\alpha=0.952$, moderate renormalization $\alpha = 0.8$, and stronger renormalization $\alpha = 0.5$, respectively. Note that the Fermi surfaces are more sensitive near the Brillouin zone boundaries. }\label{fig:fermisurface}
\end{figure}

The fermion part of the action consists of two decoupled spins of fermions $s=\pm$ ($s=\uparrow,\downarrow$ in some subsequent notations):
\begin{eqnarray}
Z_{\chi}\left[-b\right]&=&\int D\bar{\chi}D\chi e^{-\mathcal{L}
_{\chi}[b]}\nonumber\\
-\mathcal{L}
_{\chi}\left[b\right]&=&\sum_{sn\mu}\left(\bar{\chi}_{s,n}\frac{\eta_\mu^s\sigma^{\mu}-1}{2}e^{-isb_{n\mu}}\chi_{s,n+\hat{\mu}}\right.\nonumber\\\nonumber&+&\left.\bar{\chi}_{s,n+\hat{\mu}}\frac{-\eta_\mu^s\sigma^{\mu}-1}{2}e^{isb_{n\mu}}\chi_{s,n}\right)\nonumber\\
&+&\sum_{sn}\left(M+\mu\sigma^z\right)\bar{\chi}_{s,n}\chi_{s,n}\label{eq:caction}
\end{eqnarray}
where $z$ is the imaginary-time direction. $M$ controls the mass of Dirac fermions, and $\mu$ is the chemical potential. $\sigma$ labels an orbital degree of freedom for two bands separated by a gap of size $2M-6$, and $\eta^s_y=s$ and $\eta^s_x=\eta^s_z=1$ are present to preserve the time-reversal symmetry. We choose this particular fermion action so that the theory will take a special simple form after integrating out the boson and gauge fields \footnote{The choice can be generalized by switching $\sigma^x$, $\sigma^z$ and $\eta^s_y\sigma^y$ and adding factors to the hopping amplitudes.}, which is vital for our later analysis. For clarity, the two-dimensional Hamiltonian for the spin-up components of Eq. \ref{eq:caction} is:
\begin{eqnarray}
H_{2D\uparrow}&=&\sum_k c^\dagger_k c_k \times \left[\sigma^z\left(M-1-\cos(k_x)-\cos(k_y)\right)\right.  \nonumber\\
&+&\left.\sigma^x \sin(k_y) - \sigma^y \sin(k_x)+\mu\right]
\label{eq:cham}
\end{eqnarray}
where $k$ is the single-particle momentum of the electrons. The spin-down Hamiltonian is its time reversal counterpart. Without the coupling to $b_{n\mu}$, the Fermi surfaces for $M=4$ and $\mu=3.33$, $M'=5$ and $\mu'=4.17$, as well as $M"=8$ and $\mu"=6.67$ are illustrated in Fig.\ref{fig:fermisurface}. $M$ and $\mu$ are our major tuning parameters on the fermion dynamics.

The overall model in Eq. \ref{eq:totalaction} describes the above fermions and bosons coupled via an intermediate gauge field. The gauge field imposes the local constraints $n_{\chi_\uparrow} - n_{\chi_\downarrow} + n_\theta$, so the bosons are effectively coupled to the fermion spin degrees of freedom. We will integrate out the boson and gauge fields, and the boson parameters $\alpha$ and $\beta$ will appear in the renormalized fermion hopping amplitudes as well as newly generated interactions. In addition, we can generalize the initial fermion action $Z_\chi$ to that of a weak-interacting Fermi liquid without invalidating the Fermi surface. The model has global $U(1)$ charge conservation and $U(1)$ spin symmetry (along $s_z$).

We also note that $Z_\chi\left[-b\right]$ is time-reversal symmetric - the complex phase of the spin-up electrons cancels that of the spin down electrons and the boson partition function $Z_\theta(b)$ is also positive definite. Therefore, the overall system $Z$ can be calculated with Monte Carlo algorithm without the notorious sign problem\footnote{Another sign-problem-free scenario is to make the $\mathcal{L}_\chi$ real, e.g., model with only $\sigma_x$ and $\sigma_z$ and couplings to $\mathbb Z_2$ phases.}. However, instead of studying the phases corresponding to various parameters $\alpha$, $\beta$, $M$ and $\mu$ via numerical simulations, here we take an analytical approach towards a schematic phase diagram, and importantly, the conditions and microscopic mechanism for the various underlying phases. Nevertheless, the numerical approach will be helpful for verification purposes that we plan to study in a subsequent paper.

\emph{Effective interacting fermion model-- }Let's consider a fermion-only action $\mathcal{L}
_\chi'=\mathcal{L}
_\chi+\mathcal{L}
_{int}$ with $\mathcal{L}
_\chi$ in Eq. \ref{eq:caction} and:
\begin{eqnarray}
-\mathcal{L}
_{int}&=&\sum_{n\mu}B\left[L_{\uparrow}R_{\downarrow}+R_{\uparrow}L_{\downarrow}\right] \nonumber\\
&+&C\left[L_{\uparrow}R_{\uparrow}+L_{\uparrow}L_{\downarrow}+R_{\downarrow}R_{\uparrow}+L_{\downarrow}R_{\downarrow}\right] \nonumber\\
&+&D\left[L_{\uparrow}R_{\uparrow}L_{\downarrow}+L_{\uparrow}L_{\downarrow}R_{\downarrow}+R_{\uparrow}L_{\downarrow}R_{\downarrow}+L_{\uparrow}R_{\uparrow}R_{\downarrow}\right] \nonumber\\
&+&E\left[L_{\uparrow}R_{\uparrow}L_{\downarrow}R_{\downarrow}\right]
\label{eq:interactions}
\end{eqnarray}
where $B$, $C$, $D$ and $E$ are interaction strengths, and
\begin{eqnarray}
L_{\uparrow}&=&\bar{\chi}_{\uparrow,n}\frac{\sigma^{\mu}-1}{2}\chi_{\uparrow,n+\hat{\mu}}\nonumber\\
L_{\downarrow}&=&\bar{\chi}_{\downarrow,n}\frac{\eta^{-}_\mu\sigma^{\mu}-1}{2}\chi_{\downarrow,n+\hat{\mu}}\nonumber\\
R_{\uparrow}&=&\bar{\chi}_{\uparrow,n+\hat{\mu}}\frac{-\sigma^{\mu}-1}{2}\chi_{\uparrow,n}\nonumber\\
R_{\downarrow}&=&\bar{\chi}_{\downarrow,n+\hat{\mu}}\frac{-\eta^{-}_{\mu}\sigma^{\mu}-1}{2}\chi_{\downarrow,n}
\end{eqnarray}
are short-hand notations for the quadratic hopping terms between sites $n$ and $n+\hat\mu$ that are also present in $\mathcal{L}
_\chi$.

Grassmann algebra allows us to expand $Z'_\chi$ exactly
\begin{eqnarray}Z'_\chi=\int D\bar{\chi}D\chi \prod_{sn} e^{\bar\chi_{s,n} (M+\mu\sigma^z) \chi_{s,n}}\prod_{n\mu} Z_{n\mu}\end{eqnarray}
to each link:
\begin{eqnarray}
Z_{n\mu}&=&1+R_{\uparrow}+L_{\uparrow}+R_{\downarrow}+L_{\downarrow}+\left(B+1\right)\left[L_{\uparrow}R_{\downarrow}+R_{\uparrow}L_{\downarrow}\right]\nonumber\\
&+&\left(C+1\right)\left[L_{\uparrow}R_{\uparrow}+L_{\uparrow}L_{\downarrow}+R_{\downarrow}R_{\uparrow}+L_{\downarrow}R_{\downarrow}\right]
\nonumber\\
&+&\left(D+2C+B+1\right)\label{eq:expansion}\\&\times&\left[L_{\uparrow}R_{\uparrow}L_{\downarrow}+L_{\uparrow}L_{\downarrow}R_{\downarrow}+R_{\uparrow}L_{\downarrow}R_{\downarrow}+L_{\uparrow}R_{\uparrow}R_{\downarrow}\right]\nonumber\\
&+&\left(E+4D+2C^{2}+B^{2}+4C+2B+1\right)\left[L_{\uparrow}R_{\uparrow}L_{\downarrow}R_{\downarrow}\right]\nonumber
\end{eqnarray}
Note that here the derivation is exact, and we only have a limited number of terms since those containing $L_{\uparrow}L_{\uparrow}=0$ and so on vanish by construction\cite{Jingyuan2017}.

Now we look at the initial theory $Z$ in Eq. \ref{eq:totalaction} and integrate out the boson and gauge fields\cite{Jingyuan2017}. Although the initial fermions themselves are non-interacting ($\mathcal{L}_{int}=0$) or weakly-interacting ($|B|,|C|\ll 1$, $D, E=0$) to start with, the coupling to the gauge and boson fields introduces nontrivial renormalization and interactions. The consequence is to give a factor $1$ to each of the terms in Eq. \ref{eq:expansion} with total current difference $\Delta j_{n\mu}=0$ between the spin up and spin down electrons such as $L_{\uparrow}R_{\uparrow}$ and $L_{\uparrow}L_{\downarrow}$, a factor $\alpha$ to each of the terms with total current difference $\Delta j_{n\mu}=\pm1$ such as $R_{\uparrow}$ and $L_{\downarrow}$, and a factor $\beta$ to each of the terms with total current difference $\Delta j_{n\mu}=\pm2$ such as $L_\uparrow R_{\downarrow}$. After renormalizing $\bar\chi$ and $\chi$ by a factor of $\left|\alpha\right|^{-1/2}$ to make the quadratic term invariant, we map the initial $Z$ ($B=C=0$) to that of a purely fermionic $\mathcal{L}'_\chi$ exactly and analytically with the following coefficients:
\begin{eqnarray}
M'&=&M/\alpha\nonumber\\
\mu'&=&\mu/\alpha\nonumber\\
B'&=&\beta/\alpha^{2}-1\nonumber\\
C'&=&1/\alpha^{2}-1\label{eq:newinteractions}\\
D'&=& 2-(\beta+1)/\alpha^2 \nonumber\\
E'&=& -(\beta^2+1)/\alpha^4+4(\beta+1)/\alpha^2 -6\nonumber
\end{eqnarray}
Generalizations to allow interactions between the initial fermion system is also straightforward. For example, we instead have $B'=\left(1+B\right)\beta/\alpha^{2}-1$ for the $L_{\uparrow}R_{\downarrow}+R_{\uparrow}L_{\downarrow}$ term in the effective theory if the initial $B$ is finite. We emphasize that the derivation is exact up to this step, and we have mapped the original strong-coupling theory of bosons and fermions to an \emph{effective} fermion-only theory with parameters in Eq. \ref{eq:newinteractions} - a new set of interactions, as well as the mass and the chemical potential for the quadratic terms.

\begin{figure}
\includegraphics[scale=0.45]{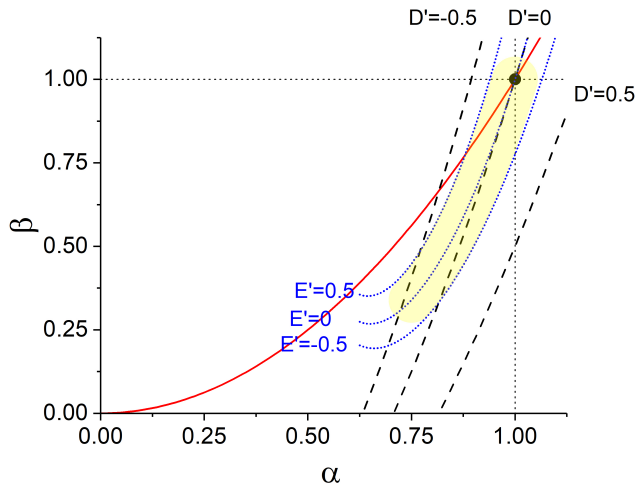}
\caption{The yellow banana-shaped area schematically denotes the $(\alpha,\beta)$ boson parameter space where the resulting effective theory is weak-coupling. There, all coupling strengths $|B'|,|C'|,|D'|,|E'|\ll 1$ starting from non-interacting fermions ($B=C=D=E=0$) coupled to the bosons. The $(\alpha,\beta)=(1,1)$ point is where the resulting effective theory is still non-interacting. The parabola in red is $\beta=\alpha^2$ and separates regions with $B'>0$ and $B'<0$.}\label{fig:wcregion}
\end{figure}

The quadratic terms have their usual band theory interpretation. In Fig. \ref{fig:wcregion}, we illustrate the parameter regions for $(\alpha,\beta)$ where all the attained interactions $|B'|,|C'|,|D'|,|E'|\ll 1$ starting from non-interacting fermions coupled to bosons. We are delighted to see that there is a reasonably large overlapped region where all effective interactions are weak in comparison with the Fermi energy, combined with the fact that the interaction terms involve higher-order operators, allows us to interpret the low-energy physics of the effective theory in the weak-coupling framework - the Fermi liquid theory, and in turn, give a controlled analysis about the resulting phases of the original strongly-correlated systems.

\emph{Fermi liquid, non-Fermi liquid, or Mott insulator?-- } We first focus on the quadratic terms. There the impacts by the bosons are reflected on the resulting Fermi surface, as $M'$ and $\mu'$ explicitly depend on $\alpha$ in Eq. \ref{eq:newinteractions}. There are three scenarios:

\begin{figure}
\includegraphics[scale=0.45]{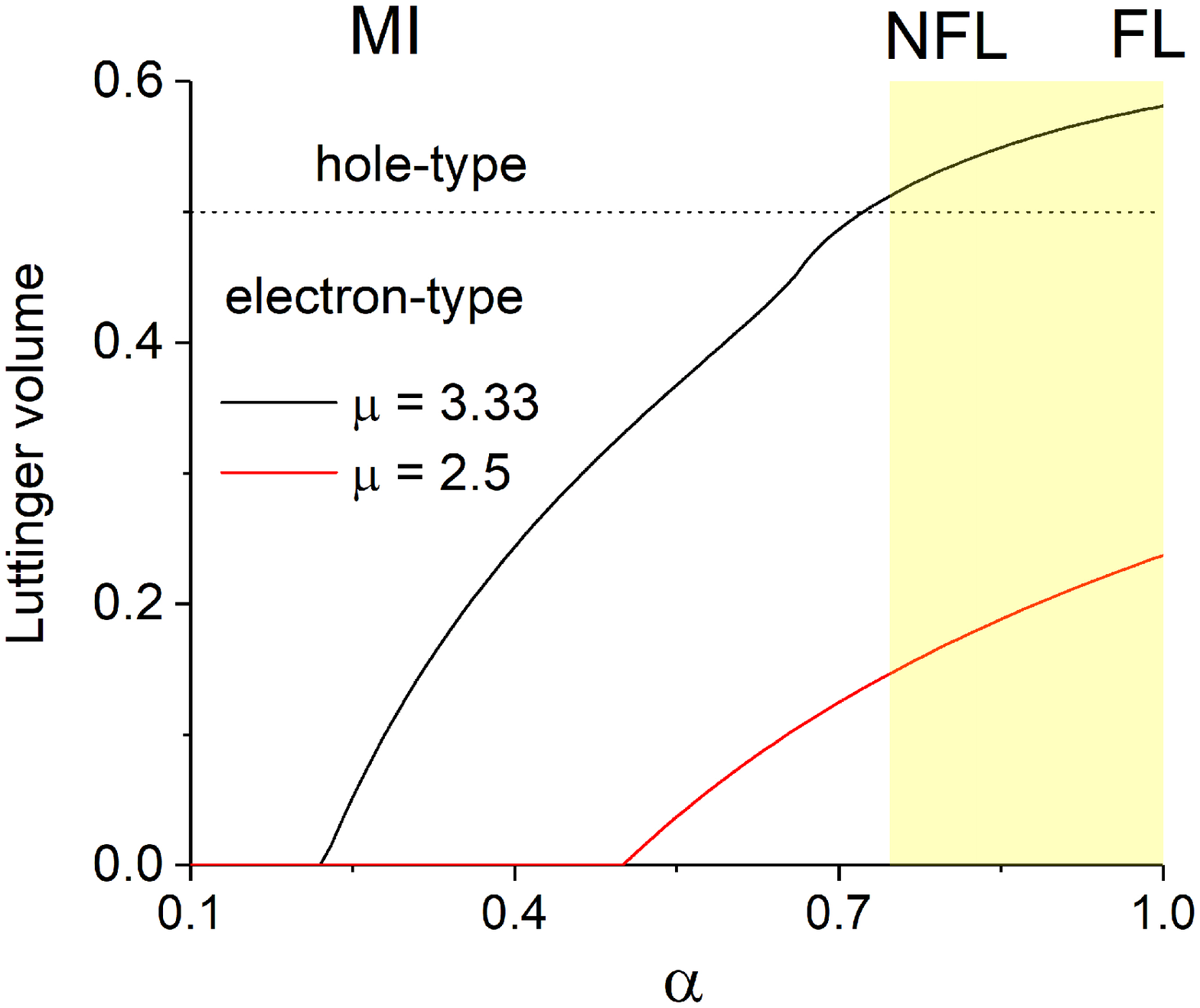}
\includegraphics[scale=0.45]{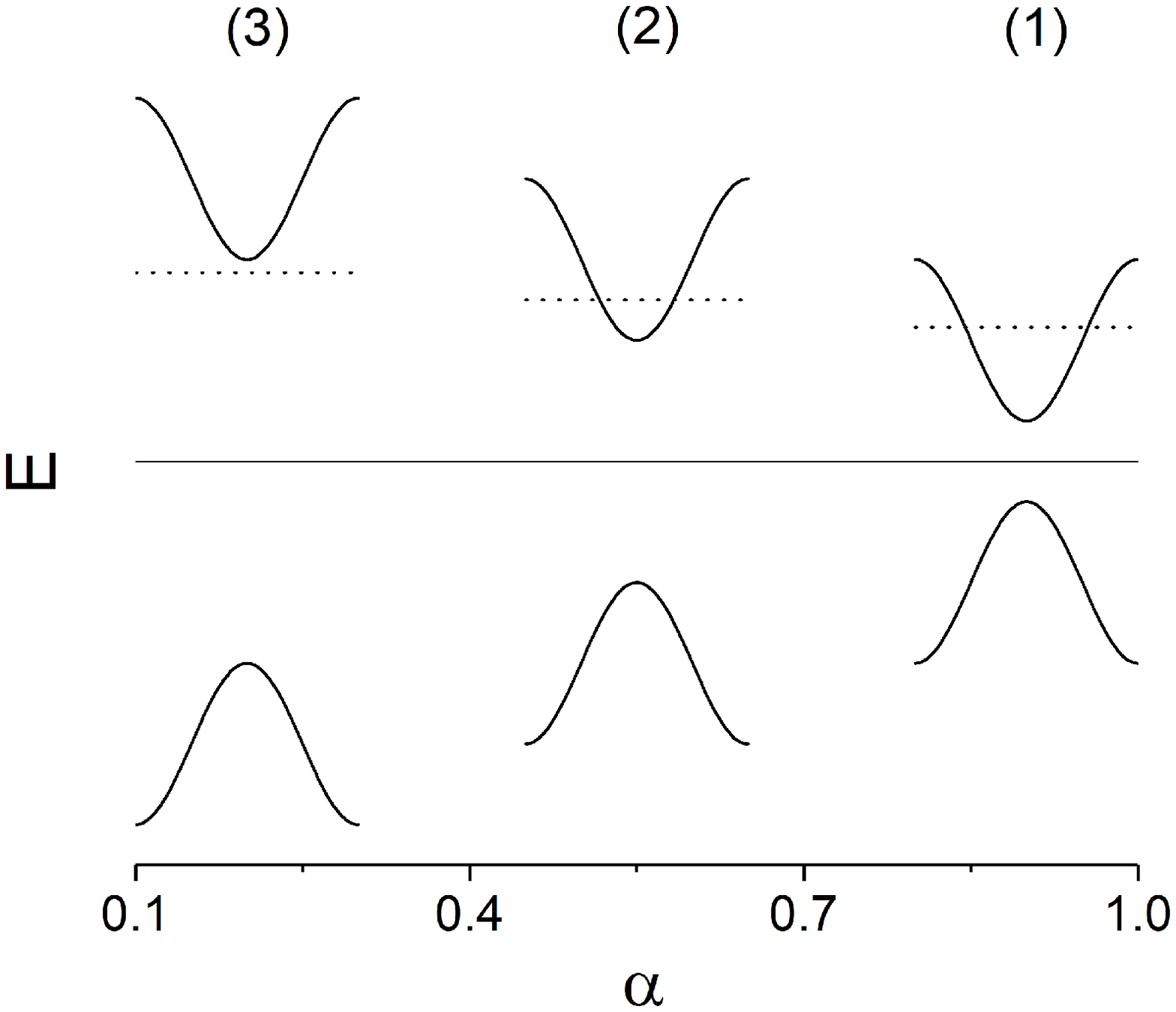}
\caption{Upper: the Luttinger volumes of the emergent Fermi surfaces at various boson parameter $\alpha$. The black and red curves are for the initial chemical potential is $\mu=3.33$ and $\mu=2.5$, respectively. $M=4$ for both. The limit $\alpha=1$ corresponds to the original Fermi surface without boson intervention. The Fermi surface undergoes a Lifshitz transition across the horizontal dotted line, and the carrier type changes sign. The yellow region is where we have weak-coupling interpretation in the effective theory, see Fig. \ref{fig:wcregion}. Lower: schematic change of the chemical potential and band gap for the quadratic terms in the effective theory.}\label{fig:Lv}
\end{figure}

(1) $\alpha\sim 1$, therefore $M'\sim M$, $\mu' \sim \mu$, and the Fermi surface is minimally impacted, see the cross symbols in Fig. \ref{fig:fermisurface} for $\alpha = 0.952$ as compared to the pristine Fermi surface in black. The resulting state remains approximately a Fermi liquid.

(2) $\alpha<1$, therefore a moderate deviation in $M'$ and $\mu'$ and the corresponding Fermi surface is observed, see the blue and red Fermi surfaces in Fig. \ref{fig:fermisurface} for $\alpha=0.8$ and $\alpha=0.5$, respectively. The weak-coupling region in yellow in Fig. \ref{fig:wcregion} and \ref{fig:Lv} at relatively large $\alpha$ allow us to establish an emergent Fermi liquid unambiguously. Essentially, the Luttinger volume of the system is noticeably different from that in the original fermion theory, see Fig. \ref{fig:Lv}, indicating an explicit violation of the Luttinger theorem. On the other hand, even at relatively smaller $\alpha$ and the resulting strong coupling in the effective theory forbids a simple Fermi liquid interpretation, it is unlikely that a double-violation of the Luttinger theorem would restore the Luttinger volume to match the very original fermion degrees of freedom. With an adequate amount of renormalization $\alpha=M-\mu$, the Luttinger volume passes 0.5, and the system goes across a Lifshitz transition, signaling an interesting carrier type change. Strikingly, even if the chemical potential $\mu$ is slightly above the band top at $M+1$ and an initial Fermi surface is absent, the coupling to bosons parameterized by $\alpha<1$ can help to bring the effective chemical potential $\mu'$  back into the renormalized conducting band and give rise to an emergent Fermi surface.

(3) $\alpha\ll 1$, the Fermi surface given by the quadratic terms in the effective theory completely disappears, because the chemical potential $\mu'$ falls below the band bottom $M'-3$ when $\alpha<(M-\mu)/3$. The vanishing Luttinger volume, see Fig. \ref{fig:Lv}, signals the loss of low-energy fermion degrees of freedom, leading to an interaction-driving Mott insulator even though the initial filling violates the requirement for band insulators. Starting from a Fermi liquid, this scenario complements the conventional picture by forbidding double occupancy in a single-occupancy state.

Simultaneously, the bosons should also become localized. We now have a fixed number of Schwinger bosons on each lattice sites, suggesting the fermion spins can spontaneously form a coexisting magnetic order in the Mott insulating phase. Since we cannot exclude the possibility that the net moment is zero on each site, the actual existence and type of magnetic order need further numerical investigation.

\begin{figure}
\includegraphics[scale=0.45]{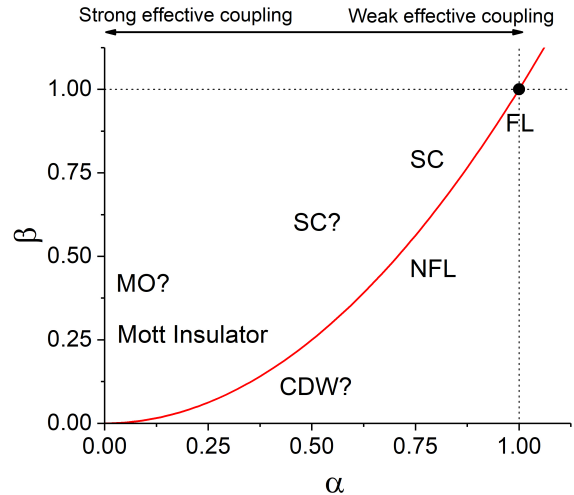}
\caption{The schematic phase diagram of the initial model in Eq. \ref{eq:totalaction} describing a finite Fermi surface of spinful electrons coupled to bosons with parameters $\alpha$ and $\beta$. The parabola in red is $\beta=\alpha^2$ separating the Fermi liquid (FL), non-Fermi liquid (NFL) with the superconductivity (SC) in the weak effective coupling region. The strong-coupling superconductivity, charge density wave (CDW), and magnetic order (MO) coexisting with the Mott insulator may or may not happen and demand further numerical investigation, hence the question marks. }\label{fig:schemepd}
\end{figure}

\emph{Conditions for emergent superconductivity-- } At weak coupling, the only instability that the emergent Fermi surface is subjected to is the superconducting pairing instability. Note that the interaction $L_{\uparrow}R_{\downarrow}+R_{\uparrow}L_{\downarrow}$ moves the electrons of opposite spins in opposite directions: $B' c^{\dagger}_{n+\hat \mu,\uparrow} c_{n, \uparrow} c^{\dagger}_{n,\downarrow} c_{n+\hat \mu, \downarrow} $, the interaction becomes attraction-like thus relevant when its coefficient $B'>0$. Equivalently, $\beta>\alpha^2/(B+1)$. If we start with non-interacting fermions ($B=0$), the allowed parameter space is above the parabola $\beta=\alpha^2$ shown as the red curves in Fig. \ref{fig:wcregion} and \ref{fig:schemepd}. If we start with a system with $B\gtrsim 0$, such intrinsic weakly-attractive interaction will be greatly enhanced $B'\gg B$, consistent with our expectation. On the other hand, even if we start with a Fermi liquid with weakly-repulsive interaction $B\lesssim 0$, if $\alpha$ is small enough, the interaction $B'>0$ can be renormalized to the attractive side. Also, give the nearest-neighbor interactions, a $d$-wave symmetric superconducting gap is preferable \cite{SU2007, Loder2010, Loder2011, Calegari2011}.

To maximize the superconducting transition temperature Tc, we need to fine-tune the strength of the pairing interaction $B'$ as well as the emergent Fermi surface determined by $M'$ and $\mu'$. Numerical studies on attractive-$U$ Hubbard model\cite{Scalettar1989} has suggested that the optimal Tc equal to a couple of percent of the bandwidth ($Tc\sim 400K$ for $W\sim 2eV$), is attained when (1) the quadratic terms give a Fermi surface with an average filling of around $1/3$, and (2) the attractive $\left|U\right|$ approaches the bandwidth $W$, yet decreases upon further increasing of $\left|U\right|$. We emphasize that the filling of the emergent Fermi surface corresponding to $M'$ and $\mu'$ can differ from the initial carrier density, especially at the stronger effective coupling. The correspondence can be derived using Eq. \ref{eq:newinteractions} given $\alpha$. In order to achieve optimal pairing interactions, since the bandwidth of our fermion model remains unchanged under renormalization, we estimate the optimal nearest neighbor pairing coefficient $B'\gtrsim 1$, which requires $\alpha \lesssim 0.7$ for $\beta\sim 1$, or $\alpha \lesssim 0.5$ for $\beta\sim 0.5$. The resulting schematic phase diagram is illustrated in Fig. \ref{fig:schemepd}. Unfortunately, this is already at strong coupling, and we cannot simply neglect the other interactions in Eq. \ref{eq:newinteractions}. Still, we expect the conclusions valid for the weak-coupling region to hold for a larger area. The actual Tc and the existence of the strong-coupling superconductivity still await numerical investigations.

Before closing, we conjecture CDW as another interesting possibility at strong coupling at relatively small $\alpha$, see Fig. \ref{fig:schemepd}. There the nearest-neighbor four-fermion interactions such as $c^{\dagger}_{n+\hat \mu,\uparrow} c_{n+\hat \mu, \downarrow} c^{\dagger}_{n,\downarrow} c_{n, \uparrow}$ and $c^{\dagger}_{n+\hat \mu,\uparrow} c_{n+\hat \mu, \uparrow} c^{\dagger}_{n,\uparrow} c_{n, \uparrow}$ become large and even comparable to the Fermi energy. This is more likely to happen below the $\beta=\alpha^2$ parabola where superconductivity is not an option, or superconductivity is suppressed by, say, an external magnetic field. Consequently, a CDW state can emerge as a result of strong coupling, irrelevant to the Fermi surface nesting scenario necessary in the weak coupling limit \footnote{The weak-coupling CDW due to the perfect nesting of the emergent Fermi surface at the Lifshitz transition is an artifact of the nearest neighbor hopping and not generic.}.

\emph{Summary and discussions-- }We have studied a model of strongly-correlated systems where a Fermi surface of spinful electrons couples to bosons via an intermediate gauge field. The gauge field constraints the boson occupancy to the fermion spin, allowing us to interpret the bosons as the Schwinger bosons of the magnetic order. By integrating out the boson and gauge fields, we obtain an effective fermion model that can become weak-coupling in chosen regions of the initial parameters. In particular, we find an emergent Fermi liquid in the effective theory that has a different Luttinger volume from the initial fermions, therefore violates the Luttinger theorem, and signals non-Fermi liquid phenomena. The emergent Fermi surface becomes unstable towards $d$-wave superconductivity if the effective theory also generates relevant pairing interactions from the boson couplings. Strong renormalization by the bosons is favorable towards attractive pairing interactions, but it also impacts the emergent Fermi surface adversely. These two effects have to be balanced for optimal superconductivity. We also interpret Mott insulator, magnetic order, charge density wave, and strong-coupling superconductivity in the same framework yet at strong effective coupling.

In addition to the various interesting phases adjacent to superconductivity, we also observe in this model unusual phenomena: (1) the emergent Fermi surface describing the collective quasi-particle excitations may have a much smaller Luttinger volume than the expected value according to the electron density; (2) upon tuning, the emergent Fermi surface may undergo a Van Hove singularities, and the carrier type may change sign; (3) the emergent Fermi surface sees most apparent changes near the Brillouin zone boundaries, see Fig. \ref{fig:fermisurface}, suggesting the electrons are most sensitive towards the boson coupling there; (4) optimal superconductivity is not immediately associated with the disappearance of magnetic order, but rather separated from it by a finite distance in the phase space, since it requires a finite volume of the emergent Fermi surface; (5) both $M'$ and $\mu'$ are much larger than their non-interacting values $M$ and $\mu$ at strong renormalization (small $\alpha$), therefore the entire band is lifted to higher energies, see Fig. \ref{fig:Lv}. A large-gap substrate that does not interfere with the materials when the influences of the boson impacts are weak may counter-intuitively act as a charge reservoir and influence $\mu'$ when the bosons become non-negligible. Interpreting preexisting experiments on realistic materials in our framework is important yet subjective and controversial, and we leave such discussions to private communications. Before ending, we note that the last effect may offer another tuning parameter in search for both the optimal effective interactions and optimal emergent Fermi surface volume for the enhanced superconducting transition temperature.

\emph{Acknowledgement-- } YZ acknowledges insightful discussions with Ashvin Vishwanath, Jing-Yuan Chen, Aaron Hui, and especially, Eun-Ah Kim, Brad Ramshaw and Steven A. Kivelson. YZ is supported by Bethe postdoctoral fellowship at Cornell University, W.M. Keck Foundation, and DOE support under award DE-SC001313. In addition, YZ is grateful to the hospitality of KITP, and supported in part by the National Science Foundation under Grant No. NSF PHY-1125915.

\bibliographystyle{apsrev4-1}
\bibliography{refs}

\end{document}